\documentclass[reprint,amsmath,amssymb,aps,]{revtex4-2}

\usepackage{graphicx}
\usepackage{dcolumn}
\usepackage{bm}

\usepackage[]{hyperref}
\usepackage{color} % color
\usepackage{amsmath}

\begin{document}

\title{Frequency-tunable biphoton generation via spontaneous four-wave mixing}

\author{Jiun-Shiuan Shiu,$^{1,2}$ Chang-Wei Lin,$^{1}$ Yu-Chiao Huang,$^{1}$ Meng-Jung Lin,$^{1}$ I-Chia Huang,$^{1}$ Ting-Ho Wu,$^{1}$ Pei-Chen Kuan,$^{1}$ and Yong-Fan Chen$^{1,2,}$}

\email{yfchen@mail.ncku.edu.tw}

\affiliation{
$^1$Department of Physics, National Cheng Kung University, Tainan 70101, Taiwan\\ 
$^2$Center for Quantum Frontiers of Research $\&$ Technology, Tainan 70101, Taiwan
}

%\date{\today}

%%%%%%%%%%%%%%%%%%%%%%%%%%%%%%%%%%%%%%%%%%%%%%%%%%%%%%%%%%%%%%%%%%%%%%%%%%%%%%%%%%%%%%%%%%%%%%%%%%%%%
%%%%%%%%%%%%%%%%%%%%%%%%%%%%%%%%%%%%%%%%%%%%%%%%%%%%%%%%%%%%%%%%%%%%%%%%%%%%%%%%%%%%%%%%%%%%%%%%%%%%%

\begin{abstract}

We present experimental results on tuning biphoton frequency by introducing a detuned coupling field in spontaneous four-wave mixing (SFWM), and examine its impact on the pairing ratio. This tunability is achieved by manipulating the inherent electromagnetically induced transparency (EIT) effect in the double-$\Lambda$ scheme. Introducing a detuned coupling field degrades the efficiency of EIT-based stimulated four-wave mixing, which in turn reduces the biphoton pairing ratio. However, this reduction can be mitigated by increasing the optical power of the coupling field. Additionally, we observe that blue- and red-detuning the biphoton frequency results in distinct temporal profiles of biphoton wavepackets due to phase mismatch. These findings provide insights into the mechanisms of frequency-tunable biphoton generation via SFWM, and suggest potential optimizations for applications in quantum communication and information processing.

\end{abstract}

%%%%%%%%%%%%%%%%%%%%%%%%%%%%%%%%%%%%%%%%%%%%%%%%%%%%%%%%%%%%%%%%%%%%%%%%%%%%%%%%%%%%%%%%%%%%%%%%%%%%%
%%%%%%%%%%%%%%%%%%%%%%%%%%%%%%%%%%%%%%%%%%%%%%%%%%%%%%%%%%%%%%%%%%%%%%%%%%%%%%%%%%%%%%%%%%%%%%%%%%%%%

\maketitle

%%%%%%%%%%%%%%%%%%%%%%%%%%%%%%%%%%%%%%%%%%%%%%%%%%%%%%%%%%%%%%%%%%%%%%%%%%%%%%%%%%%%%%%%%%%%%%%%%%%%%
%%%%%%%%%%%%%%%%%%%%%%%%%%%%%%%%%%%%%%%%%%%%%%%%%%%%%%%%%%%%%%%%%%%%%%%%%%%%%%%%%%%%%%%%%%%%%%%%%%%%%

\newcommand{\FigOne}{
    \begin{figure}[t]
    \centering
    \includegraphics[width = 8.5 cm]{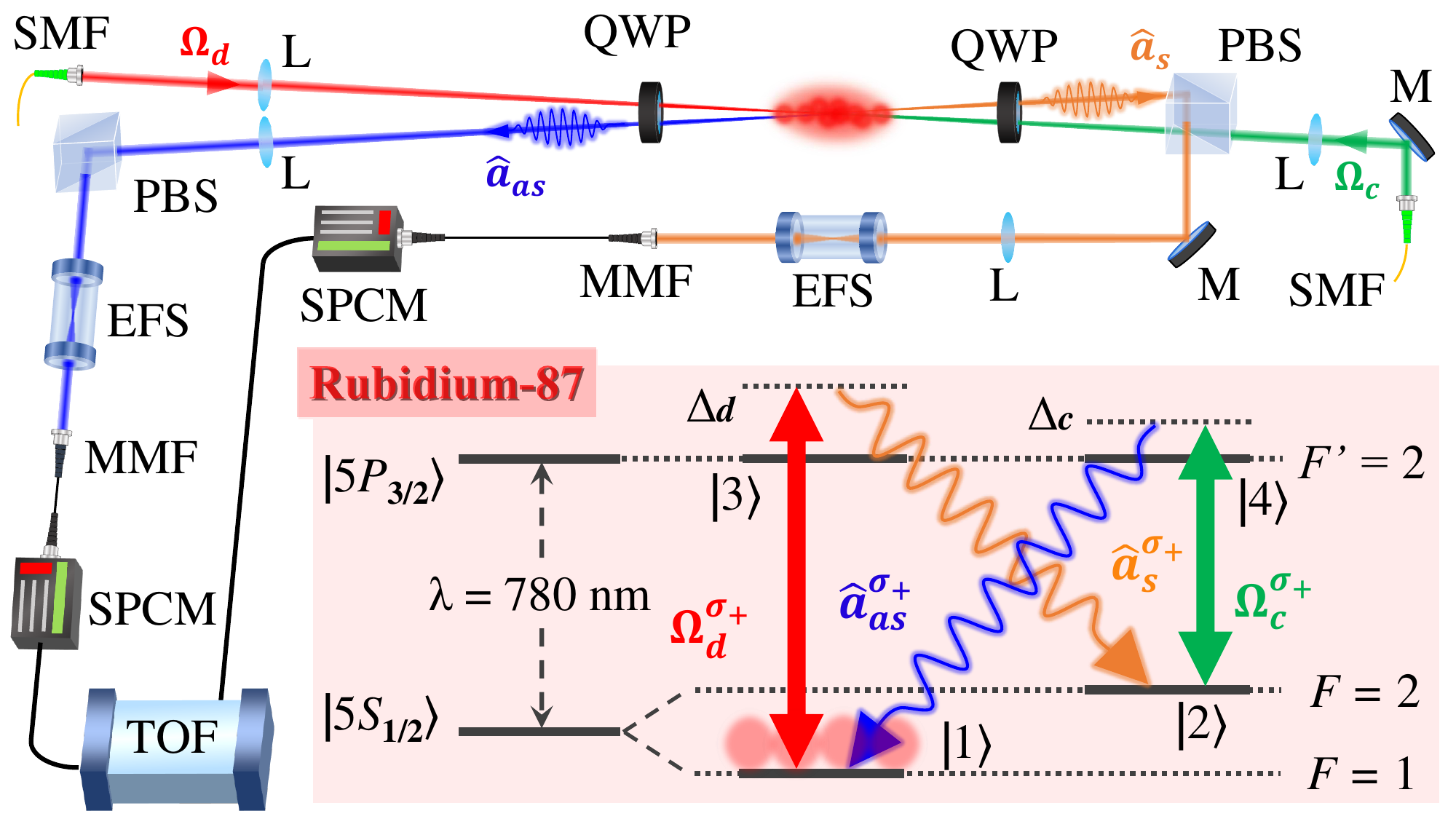}
    \caption{
The diagram illustrates the double-$\Lambda$ SFWM system and experimental setup, including mirrors (M), lenses (L), polarizing beam splitters (PBS), quarter-wave plates (QWP), single-mode fibers (SMF), multi-mode fibers (MMF), etalon filter sets (EFS), single-photon counting modules (SPCM), and a time-of-flight multiscaler (TOF). The inset shows the energy levels of the $^{87}$Rb atoms involved.
}
    \label{fig:setup}
    \end{figure}
}

\newcommand{\FigTwo}{
    \begin{figure}[t]
    \centering
    \includegraphics[width = 9.0 cm]{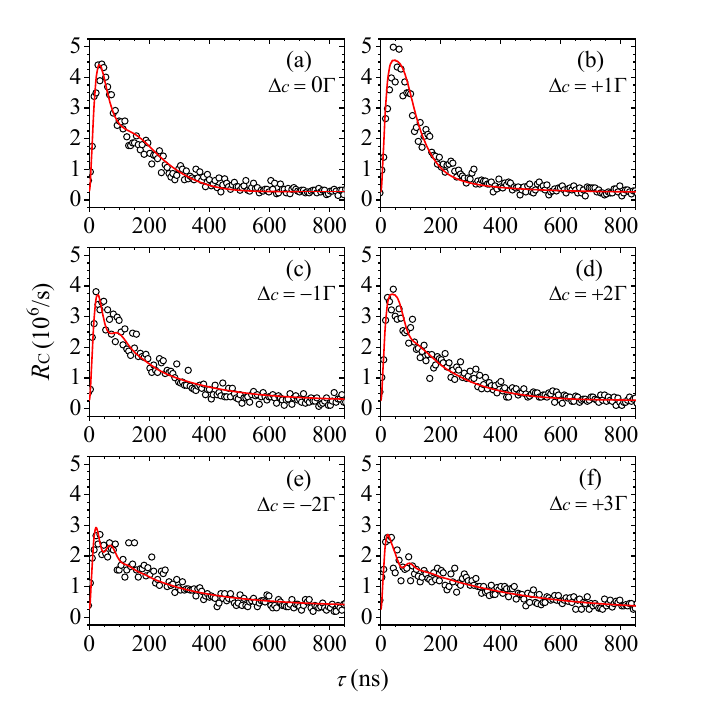}
    \caption{
The asymmetry in biphoton wavepackets under the slow light regime is observed by tuning the coupling field frequency to achieve blue- and red-detuning $\Delta_c$. Experimental data (circles) are compared with theoretical predictions (red solid curves). The parameters used are OD = 10, $\Omega_d=1\Gamma$, $\Omega_c=1\Gamma$, $\Delta_d=10\Gamma$, $\gamma_{21}=0.001\Gamma$, $\Delta kL=0.37\pi$, and $\Delta\tau=6.4$ ns. Subplots (a) to (f) show (a) $\Delta_c=0\Gamma$, (b) $\Delta_c=+1\Gamma$, (c) $\Delta_c=-1\Gamma$, (d) $\Delta_c=+2\Gamma$, (e) $\Delta_c=-2\Gamma$, and (f) $\Delta_c=+3\Gamma$.
}
    \label{fig:slowlight}
    \end{figure}
}

\newcommand{\FigThree}{
    \begin{figure}[t]
    \centering
    \includegraphics[width = 8.8 cm]{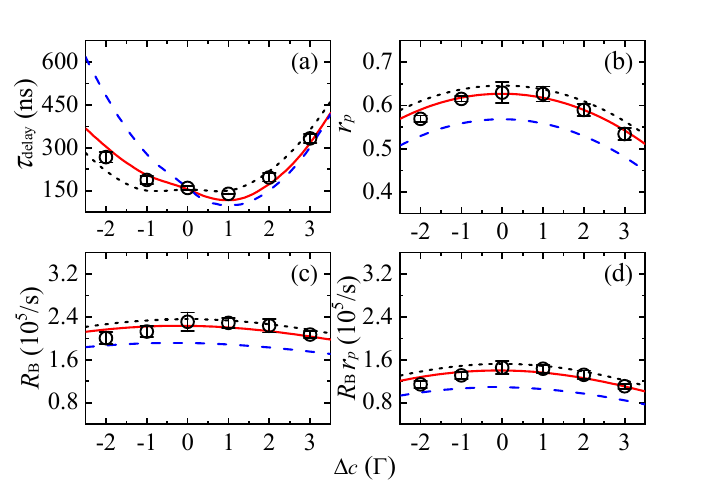}
    \caption{
The results for frequency-tunable biphotons under the slow light regime are presented, including the wavepacket delay time, pairing ratio, and generation rate. The hollow circles represent experimental data for $\Delta kL=0.37\pi$, while the red solid curves correspond to theoretical predictions. The black dotted and blue dashed curves represent theoretical scenarios for $\Delta kL=0$ and $\Delta kL=0.74\pi$, respectively. Other parameters are OD = 10, $\Omega_d=1\Gamma$, $\Omega_c=1\Gamma$, $\Delta_d=10\Gamma$, and $\gamma_{21}=0.001\Gamma$. Subplots (a) to (d) illustrate (a) $\tau_{\rm delay}$ vs. $\Delta_c$, (b) $r_p$ vs. $\Delta_c$, (c) $R_{\rm B}$ vs. $\Delta_c$, and (d) $R_{\rm B}r_p$ vs. $\Delta_c$.
}
    \label{fig:slowlight2}
    \end{figure}
}

\newcommand{\FigFour}{
    \begin{figure}[t]
    \centering
    \includegraphics[width = 9.0 cm]{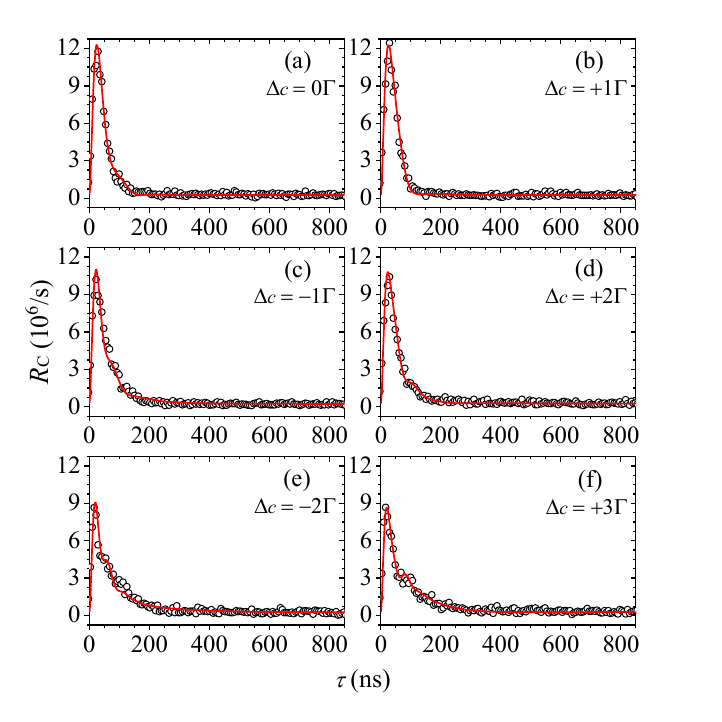}
    \caption{
The asymmetry in biphoton wavepackets under the damped Rabi oscillation regime is observed by tuning the coupling field frequency to achieve blue- and red-detuning $\Delta_c$. Experimental data (circles) are compared with theoretical predictions (red solid curves). The parameters used are OD = 10, $\Omega_d=1\Gamma$, $\Omega_c=2\Gamma$, $\Delta_d=10\Gamma$, $\gamma_{21}=0.001\Gamma$, $\Delta kL=0.37\pi$, and $\Delta\tau=6.4$ ns. Subplots (a) to (f) show (a) $\Delta_c=0\Gamma$, (b) $\Delta_c=+1\Gamma$, (c) $\Delta_c=-1\Gamma$, (d) $\Delta_c=+2\Gamma$, (e) $\Delta_c=-2\Gamma$, and (f) $\Delta_c=+3\Gamma$.
}
    \label{fig:Rabi}
    \end{figure}
}

\newcommand{\FigFive}{
    \begin{figure}[t]
    \centering
    \includegraphics[width = 8.8 cm]{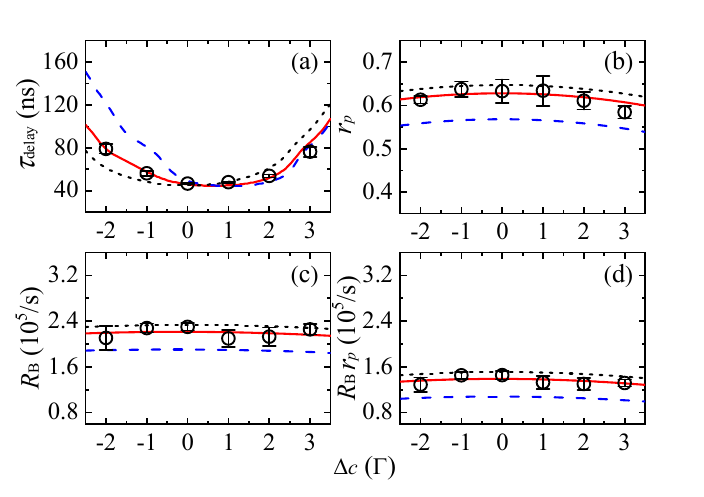}
    \caption{
The results for frequency-tunable biphotons under the damped Rabi oscillation regime are presented, including the wavepacket delay time, pairing ratio, and generation rate. The hollow circles represent experimental data for $\Delta kL=0.37\pi$, while the red solid curves correspond to theoretical predictions. The black dotted and blue dashed curves represent theoretical scenarios for $\Delta kL=0$ and $\Delta kL=0.74\pi$, respectively. Other parameters are OD = 10, $\Omega_d=1\Gamma$, $\Omega_c=2\Gamma$, $\Delta_d=10\Gamma$, and $\gamma_{21}=0.001\Gamma$. Subplots (a) to (d) illustrate (a) $\tau_{\rm delay}$ vs. $\Delta_c$, (b) $r_p$ vs. $\Delta_c$, (c) $R_{\rm B}$ vs. $\Delta_c$, and (d) $R_{\rm B}r_p$ vs. $\Delta_c$.
}
    \label{fig:Rabi2}
    \end{figure}
}

%%%%%%%%%%%%%%%%%%%%%%%%%%%%%%%%%%%%%%%%%%%%%%%%%%%%%%%%%%%%%%%%%%%%%%%%%%%%%%%%%%%%%%%%%%%%%%%%%%%%%
%%%%%%%%%%%%%%%%%%%%%%%%%%%%%%%%%%%%%%%%%%%%%%%%%%%%%%%%%%%%%%%%%%%%%%%%%%%%%%%%%%%%%%%%%%%%%%%%%%%%%

\section{Introduction} \label{sec:Introduction}

Over the past two decades, spontaneous four-wave mixing (SFWM) in atomic media has garnered considerable attention as an effective approach for generating temporally correlated biphotons, serving as heralded single-photon sources~\cite{SFWM1, SFWM2, SFWM3, SFWM4, SFWM5, SFWM6, SFWM7, SFWM8, SFWM9}. Compared to the DLCZ protocol~\cite{DLCZ1, DLCZ2, DLCZ3, DLCZ4, DLCZ5, DLCZ6, DLCZ7, DLCZ8, DLCZ9}, the SFWM scheme simultaneously utilizes two applied light fields, where the atomic medium acts as a $\chi^{(3)}$ nonlinear quantum interface, facilitating stimulated four-wave mixing (stimulated FWM). Although stimulated FWM establishes the directionality of temporally correlated photons, not all photons detected in the desired direction are actively involved in the process. Some photons are detected in the specified direction, while their counterparts scatter elsewhere. To elucidate this phenomenon, a previous theoretical study introduced the concept of the pairing ratio, quantifying the intrinsic proportion of biphotons generated through stimulated FWM relative to the total scattered photons~\cite{Kolchin}. Our recent experiment reveals that the pairing ratio decreases with increasing biphoton generation rate; however, this effect is mitigated by increasing the density of the atomic ensemble~\cite{Shiu}. These findings enhance the understanding of pairing dynamics in SFWM and underscore the broader importance of investigating the intrinsic properties of biphoton sources, paving the way for optimizing their performance across a wide range of quantum applications.

Building on these insights, the investigation of biphoton sources—including their diverse configurations and bandwidths—is essential for advancing high-performance source development and unlocking their potential in quantum applications. To date, broadband biphoton sources based on diamond-type~\cite{Diamond1} and cascade-type~\cite{Ladder1} configurations have attracted significant attention, as evidenced by recent breakthroughs in single-photon synchronization using quantum memory~\cite{Ladder2} and the automated distribution of polarization-entangled photons~\cite{Diamond2}. Broadband biphoton sources generated via spontaneous parametric down-conversion (SPDC) have also made remarkable progress in recent studies~\cite{SPDC1, SPDC2}, demonstrating their value in high-speed and frequency-multiplexed quantum applications. In parallel, narrowband biphoton sources have emerged as a complementary approach, playing an equally critical role in quantum optics and quantum information science. These sources, characterized by their elongated temporal correlation, provide the distinct advantage of long coherence times, which are essential for maintaining phase stability over extended distances. While their narrow bandwidth may increase processing time in certain quantum computing protocols, it greatly enhances compatibility with atomic systems and facilitates long-distance quantum communication, underscoring their indispensable value.

Among these narrowband biphoton sources, double-$\Lambda$ SFWM stands out for its intrinsic $\Lambda$-type electromagnetically induced transparency (EIT)~\cite{EIT1, EIT2, EIT3}, making it one of the most extensively studied schemes in this category. The inherent $\Lambda$-type EIT structure of double-$\Lambda$ SFWM is particularly notable for its low optical power requirement to achieve near-unity transmission, while also enabling the generation of photons with narrow bandwidths. This characteristic leads to long biphoton temporal correlation times in both cold~\cite{bandwidth1, bandwidth2} and hot atoms~\cite{bandwidth3, bandwidth4}. The narrowband property makes these photons highly suitable for frequency-sensitive applications, such as EIT-based quantum memories~\cite{QM1, QM2, QM3}, quantum frequency converters~\cite{QFC1, QFC2, QFC3}, and other potential quantum device implementations~\cite{device1, device2, device3}. However, achieving optimal efficiency in these frequency-sensitive systems often requires precise tuning of the incident photon frequency, highlighting the critical importance of frequency tunability—a characteristic that remains to be systematically explored in narrowband biphoton sources.

In this paper, we investigate the biphoton frequency tunability using a backward double-$\Lambda$ SFWM scheme. This backward configuration, commonly employed in cold-atom systems, effectively filters optical leakages without introducing significant phase mismatch~\cite{backward1, backward2, backward3}. Our experiments show that introducing one-photon detuning to the built-in EIT results in a decrease in both the pairing ratio and the biphoton generation rate. However, this reduction can be mitigated by increasing the applied coupling field power. Furthermore, we observe an asymmetry in the biphoton temporal profile when blue- and red-detuning the frequency of the heralded photons, a phenomenon attributable to the often previously overlooked phase-mismatch effect. Notably, while the phase-matching forward scheme~\cite{forward1, forward2} does not exhibit asymmetric temporal profiles, both the backward scheme—demonstrated in this work—and the perpendicular scheme~\cite{perp1, perp2, perp3} may lead to such asymmetry when tuning the biphoton frequency. Therefore, a clear understanding of these frequency-dependent effects is essential for optimizing biphoton source performance and enabling seamless integration with diverse quantum technologies.

The paper is organized as follows. In Sec.~\ref{sec:Setup}, we describe our experimental setup and details. In Sec.~\ref{sec:Theory}, we employ Heisenberg--Langevin operator theory to delve into the biphoton generation and introduce the concept of coincidence count rate. In Sec.~\ref{sec:Results}, we demonstrate our experimental results on the coincidence count rate of frequency-controllable biphotons. In Sec.~\ref{sec:Conclusion}, we conclude our findings from this work.

%%%%%%%%%%%%%%%%%%%%%%%%%%%%%%%%%%%%%%%%%%%%%%%%%%%%%%%%%%%%%%%%%%%%%%%%%%%%%%%%%%%%%%%%%%%%%%%%%%%%%
%%%%%%%%%%%%%%%%%%%%%%%%%%%%%%%%%%%%%%%%%%%%%%%%%%%%%%%%%%%%%%%%%%%%%%%%%%%%%%%%%%%%%%%%%%%%%%%%%%%%%

\section{Experimental Setup} \label{sec:Setup}

%%%%%%%%%%%%%%%%%%%%%%%%%%%%%
\FigOne
%%%%%%%%%%%%%%%%%%%%%%%%%%%%%

Figure~\ref{fig:setup} shows our experimental setup for generating biphotons based on SFWM in a cold atomic system and the relevant energy levels of rubidium atoms. The excited states $|3\rangle$ and $|4\rangle$ are identical, meaning that the biphotons are generated using a three-level scheme. Before the SFWM experiment, the cold $^{87}$Rb atoms are prepared in the ground state $|5S_{1/2}, F=1\rangle$ using a pump laser. The distribution of Zeeman states for $|5S_{1/2}, F=1\rangle$ is approximately 20\% for $m = -1$, 40\% for $m =0$, and 40\% for $m= +1$. After population preparation, the atomic ensemble is illuminated by a far-detuned driving field with a Rabi frequency $\Omega_d$ and a nearly resonant coupling field with a Rabi frequency $\Omega_c$. The coupling field counter-propagates with the driving field, and their synchronization is achieved through injection locking with an external cavity diode laser (not shown). More details about injection locking and the synchronization can be found in the same arrangement of Ref. \cite{QFC2}. The driving field stimulates the $\sigma^+$ transition between $|5S_{1/2}, F=1\rangle$ and $|5P_{3/2}, F'=2\rangle$ with a large detuning $\Delta_d$, significantly reducing one-photon absorption. This far-detuned driving field induces spontaneous Raman transitions, resulting in the generation of Stokes photons and the concurrent creation of ground-state coherence. Simultaneously, the coupling field, acting on the $\sigma^+$ transition between $|5S_{1/2}, F=2\rangle$ and $|5P_{3/2}, F'=2\rangle$, converts the established ground-state coherence into anti-Stokes photons, leading to the emission of anti-Stokes photons to complete the FWM process. As the generation of anti-Stokes photons satisfies the two-photon resonance condition with the coupling field, we can control the biphoton frequency by adjusting the coupling detuning $\Delta_c$.

In our SFWM experiment, biphotons are generated using a 10-$\mu$s driving pulse in each 2.5-ms cycle. During the biphoton generation process, cooling of the atoms is not maintained. The atomic ensemble is initially cooled to about 300 $\mu$K, ensuring that the atoms remain nearly stationary throughout the 10-$\mu$s biphoton generation duration. The $1/e^2$ full widths of the employed driving and coupling beams are 250 and 310 $\mu$m, respectively. Since both fields drive $\sigma^+$-transitions, the generated photon pairs also exhibit $\sigma^+$ polarizations. We collect the generated Stokes (anti-Stokes) photons at a direction 1.7$^\circ$ apart from the driving (coupling) field. For each channel, an etalon filter set (EFS) is used to eliminate the optical leakage. Each EFS consists of two etalons, with an extinction ratio of approximately 30 dB and a bandwidth of around 100 MHz. The etalons in each EFS are separated by an optical isolator. The total extinction ratios for the Stokes and anti-Stokes channels are 114 dB and 124 dB, respectively. 

Biphotons are detected using fiber-coupled single-photon counting modules (SPCM, AQRH-13-FC). Upon detection, the SPCM sends an 8-ns electrical signal, which is routed through a BNC splitter to both a digital oscilloscope (DO, RTO2014) and a time-of-flight multiscaler (TOF, MCS6A-4T8). The DO records the aggregate photon counts detected by each channel, while the TOF logs the coincident counts between the two channels. When biphotons are generated as temporally correlated pairs, they reach the SPCMs within the correlation time, resulting in a non-flat biphoton wavepacket. The TOF then generates a histogram of coincident counts based on these data points, providing insights into the biphoton source.

%%%%%%%%%%%%%%%%%%%%%%%%%%%%%%%%%%%%%%%%%%%%%%%%%%%%%%%%%%%%%%%%%%%%%%%%%%%%%%%%%%%%%%%%%%%%%%%%%%%%%
%%%%%%%%%%%%%%%%%%%%%%%%%%%%%%%%%%%%%%%%%%%%%%%%%%%%%%%%%%%%%%%%%%%%%%%%%%%%%%%%%%%%%%%%%%%%%%%%%%%%%

\section{Theoretical Model} \label{sec:Theory}

\subsection{Biphoton Generation Rate} %\label{sec:Biphoton Generation Rate}

The propagating properties of the generated biphotons can be described by the following Maxwell--Schrödinger equations:
\begin{align}
&\left(
\frac{1}{c}\frac{\partial}{\partial t}+
\frac{\partial}{\partial z}
\right)\hat{a}_s
=\frac{iNg_s^*}{c}\hat{\sigma}_{23},\label{eq:1}
\\
%%%%%%%%%%%%%%%%%%%%%%%%%%%%%%%%%%%%%%%%%%%%%%%%%%%%%%%%%%
%%%%%%%%%%%%%%%%%%%%%%%%%%%%%%%%%%%%%%%%%%%%%%%%%%%%%%%%%%
&\left(
\frac{1}{c}\frac{\partial}{\partial t}-
\frac{\partial}{\partial z}+i\Delta k
\right)\hat{a}_{as}^\dagger
=-\frac{iNg_{as}}{c}\hat{\sigma}_{41},\label{eq:2}
\end{align} 
where $N$ represents the total number of atoms, and $c$ denotes the speed of light. The coupling constant between the Stokes field and atoms is $g_s = d_{32}\sqrt{\bar{\omega}_s/2\hbar\epsilon_0V}$, where $d_{32}$ indicates the dipole moment of the $|2\rangle\leftrightarrow|3\rangle$ transition, $\bar{\omega}_s$ is the central angular frequency of the Stokes field, $\epsilon_0$ is the vacuum permittivity, and $V$ is the interaction volume. Similarly, the coupling constant $g_{as}=d_{41}\sqrt{\bar{\omega}_{as}/2\hbar\epsilon_0V}$ describes the interaction between the anti-Stokes field and atoms. The operator $\hat{a}_{s(as)}$ is the annihilation operator of the Stokes (anti-Stokes) field. In our experiment, the applied light fields propagate in opposite directions, with a phase mismatch $\Delta k=|\vec{k}_{d}-\vec{k}_{s}+\vec{k}_{c}-\vec{k}_{as}|$ of 92 $\pi$/m. Since this value is primarily determined by the ground-state energy difference of 6.835 GHz, both $\Delta_d$ and $\Delta_c$ have minimal impact on it. The collective atomic operator $\hat{\sigma}_{jk}$ can be derived using Heisenberg-Langevin operator theory~\cite{Kolchin}. By applying the Fourier transform, the Stokes and anti-Stokes field operators in the frequency domain can be expressed as follows:
\begin{align}
\begin{bmatrix}
\widetilde{a}_{s}(L,\omega)\\
\widetilde{a}_{as}^{\dagger}(0,-\omega)
\end{bmatrix}
=&
\begin{bmatrix}                                        
A & B                           
\\
C & D                            
\end{bmatrix}                                          
\begin{bmatrix}
\widetilde{a}_{s}(0,\omega)\\
\widetilde{a}_{as}^{\dagger}(L,-\omega)
\end{bmatrix}
\nonumber\\
&+
\sqrt{\frac{N}{c}}
\sum_{jk}
\int_0^Ldz'
\begin{bmatrix}
  P_{jk} \\ Q_{jk}
\end{bmatrix}
\widetilde{F}_{jk}.\label{eq:3}
\end{align}
Here, $\widetilde{F}_{jk}$ represents the frequency-domain Langevin noise, and $L$ denotes the length of the atomic ensemble. The matrix elements $A$, $B$, $C$, $D$, $P_{jk}$, and $Q_{jk}$ can be determined by considering the initial boundary conditions of the field operators and noise terms~\cite{Kolchin}. Using Eq. \eqref{eq:3}, the photon generation rate $R_{s(as)}=\frac{c}{L}\langle\hat{a}_{s(as)}^\dagger\hat{a}_{s(as)}^{_{}}\rangle$ for the Stokes (anti-Stokes) field can be calculated, leading to the following expression:
\begin{align}
R_s
=&\int\frac{d\omega}{2\pi}
\left(
|B|^2
+\sum_{jk,j'k'}\int_0^Ldz\,
P_{jk}^*\mathcal{D}_{jk^\dagger,j'k'}P_{j'k'}
\right)
\nonumber\\
%-----------------------------------
\equiv&\int\frac{d\omega}{2\pi}\widetilde{R}_s(\omega),\label{eq:4}
\\
%%%%%%%%%%%%%%%%%%%%%%%%%%%%%%%%%%%%%
R_{as}
=&\int\frac{d\omega}{2\pi}
\left(
|C|^2
+\sum_{jk,j'k'}\int_0^Ldz\,
Q_{jk}\mathcal{D}_{jk,j'k'^\dagger}Q_{j'k'}^*
\right)
\nonumber\\ 
%-----------------------------------
\equiv&\int\frac{d\omega}{2\pi}\widetilde{R}_{as}(\omega),\label{eq:5}
\end{align}
where $\widetilde{R}_{s(as)}(\omega)$ represents the spectrum of the Stokes (anti-Stokes) field. The $B$ and $C$ terms involve photons participating in the stimulated FWM process, which exhibit specific propagation direction characteristics and establish temporal correlations between Stokes and anti-Stokes photons. In contrast, the integral terms associated with the diffusion coefficients $\mathcal{D}_{jk^\dagger,j'k'}$ and $\mathcal{D}_{jk,j'k'^\dagger}$ represent photons generated by spontaneous Raman scattering that do not participate in the stimulated FWM process and exhibit an isotropic emission nature. This results in the absence of temporal correlations between Stokes and anti-Stokes photons in the specific propagation direction. Therefore, the pairing ratio $r_{p}$ can be defined as the proportion of these temporally correlated photons to the total number of generated photons. Detailed theoretical discussions are provided in the supplemental material of reference~\cite{Shiu}.

In addition to calculating the photon generation rate and pairing ratio, we can derive the following normalized second-order cross-correlation function based on the field operators obtained from Eq. \eqref{eq:3}:
\begin{align}
g_{s\text{-}as}^{(2)}(\tau)=&
1+
\frac{1}{R_sR_{as}}
\left|
\int \frac{d\omega}{2\pi}
e^{-i\omega\tau}
\left(
\vphantom{\frac{A}{B}}
B^*D
\right.
\right.
\nonumber\\
%-----------------------------------
&+\sum_{jk,j'k'}\int_0^Ldz
\left.
\left.
\vphantom{\frac{A}{B}}
P_{jk}^*\mathcal{D}_{jk^\dagger,j'k'}Q_{j'k'}\right)
\right|^2.\label{eq:6}
\end{align}
The first term on the right-hand side of Eq. (6), represented by the constant value of 1, signifies the background of coincidence counts due to the uncorrelated nature between different pairs of biphotons. In contrast, the second term reveals the temporal quantum correlations of biphoton pairs resulting from the stimulated FWM process, corresponding to the biphoton wavepacket. It is worth noting that in our SFWM system, $g_s = g_{as} \equiv g$. To align with commonly used experimental parameters, we substitute $\frac{g^2N}{c}$ with $\frac{\mathrm{OD}\Gamma}{4L}$, where OD denotes the optical depth of the atomic ensemble, and $\Gamma$ represents the spontaneous decay rate of the rubidium-87 D2 line.

%%%%%%%%%%%%%%%%%%%%%%%%%%%%%%%%%%%%%%%%%%%%%%%%%%%%%%%%%%%%%%%%%%%%%%%%%%%%%%%%%%%%%%%%%%%%%%%%%%%%%
%%%%%%%%%%%%%%%%%%%%%%%%%%%%%%%%%%%%%%%%%%%%%%%%%%%%%%%%%%%%%%%%%%%%%%%%%%%%%%%%%%%%%%%%%%%%%%%%%%%%%

\subsection{Coincidence Count Rate} %\label{sec:Coincidence Count Rate}

The coincidence count rate characterizes the generation rate of anti-Stokes photons conditioned on the generation of Stokes photons. This rate is defined as $R_{\rm C}(\tau)=R_sR_{as}g_{s\text{-}as}^{(2)}(\tau)\Delta T$, where $\Delta T$ represents the collection time interval for the generated Stokes photons. In our data analysis, we transform the histogram of coincidence counts into the coincidence count rate by setting $\Delta T=1/R_s$. This conversion normalizes the number of the Stokes photons to one, meaning only a single Stokes photon is emitted from the atomic ensemble within this time interval. With this selection, the background $R_{\rm C}(\tau\rightarrow\infty)$ and the correlated area of the biphoton wavepacket $\int[R_{\rm C}(\tau)-R_{\rm C}(\tau\rightarrow\infty)]d\tau$ correspond to $R_{as}$ and $r_{p}$, respectively. Notably, under ideal conditions, $R_s$ and $R_{as}$ are equal. In practical systems, however, $R_{as}$ is slightly smaller due to phase mismatch or ground state decoherence. As a result, the biphoton generation rate $R_{\rm B}$ depends on the value of $R_{as}$.

In our experiment, we accumulate a total of $2^{18}$ receptions in the Stokes channel, each with a purity $P_s=\eta_{s}R_{s}/(\eta_{s}R_{s}+R_{\rm noise}^{s})$, where $\eta_{s}$ indicates the collection efficiency of the Stokes photons and $R_{\rm noise}^{s}$ denotes the noise count rate in the Stokes channel. Using the collection efficiency $\eta_{as}$ and the noise count rate $R_{\rm noise}^{as}$ in the anti-Stokes channel, we can establish the relationship between $R_{\rm C}$ and the coincidence counts $N_{\rm C}$ as follows:
\begin{align}
N_{\rm C}=&
2^{18}P_s\times R_{\rm C}\eta_{as}\Delta\tau
+
2^{18}P_s\times R_{\rm noise}^{as}\Delta\tau
\nonumber\\&
+
2^{18}(1-P_{s})\times(R_{\rm noise}^{as}+R_{\rm B}\eta_{as})\Delta\tau.\label{eq:7}
\end{align}
The time bin $\Delta\tau$ for detected anti-Stokes photons aligns with the time interval between experimental data points. In Eq.~\eqref{eq:7}, the first term represents the detection of biphotons, corresponding to the simultaneous registration of both Stokes and anti-Stokes photons. The second and last terms account for environmental background, resulting from pure and impure detections in the Stokes channel, respectively. If the Stokes photons are successfully detected, the impure detection in the anti-Stokes channel results in an uncorrelated background, indicated by the second term. On the other hand, it consistently contributes to the uncorrelated background when the detection in the Stokes channel does not involve Stokes photons, as represented by the last term. Finally, the experimental biphoton coincidence count rate can be expressed as $N_{\rm C}/(2^{18}P_{s}\eta_{as}\Delta\tau)\equiv R_{\rm C}+R_{\rm env}$, where $R_{\rm env}$ represents the environmental background count rate and can be determined by experimental measurements.

%%%%%%%%%%%%%%%%%%%%%%%%%%%%%
\FigTwo
%%%%%%%%%%%%%%%%%%%%%%%%%%%%%

%%%%%%%%%%%%%%%%%%%%%%%%%%%%%
\FigThree
%%%%%%%%%%%%%%%%%%%%%%%%%%%%%

%%%%%%%%%%%%%%%%%%%%%%%%%%%%%%%%%%%%%%%%%%%%%%%%%%%%%%%%%%%%%%%%%%%%%%%%%%%%%%%%%%%%%%%%%%%%%%%%%%%%%
%%%%%%%%%%%%%%%%%%%%%%%%%%%%%%%%%%%%%%%%%%%%%%%%%%%%%%%%%%%%%%%%%%%%%%%%%%%%%%%%%%%%%%%%%%%%%%%%%%%%%

\section{Results and Discussion} \label{sec:Results}

Figure~\ref{fig:slowlight} shows biphoton wavepackets (experimental $R_{\rm C}$) for different coupling detuning $\Delta_c$ with a fixed OD of 10. The circles represent experimental data obtained from TOF measurements, with a time interval $\Delta\tau$ set to 6.4 ns. The red curves are plotted based on Eq. \eqref{eq:7}, which incorporates the environmental background. The phase-mismatch parameter $\Delta kL=0.37\pi$ is calculated based on the length $L=4$ mm of our cold atomic cloud. Note that the phase mismatch here is related only to the geometric arrangement of the light fields involved in the four-wave mixing and does not take into account the dispersion effects of the medium. By setting $\Omega_d=1\Gamma$ and $\Delta_d=10\Gamma$, we measure the biphoton generation rate $R_{\rm B}$ to be approximately $2.2\times10^5$ s$^{-1}$ under the condition of a ground-state decoherence rate $\gamma_{21}=0.001\Gamma$, which is determined through an additional $\Lambda$-type EIT experiment. Moreover, the collection efficiencies $\eta_{s}$ and $\eta_{as}$, measured from our biphoton experiment for the Stokes and anti-Stokes fields, are approximately 2\% and 1\%, respectively. For more detailed information on the transmission efficiencies of various optical components in the two channels of our biphoton experiment system, please refer to the supplemental material in Ref.~\cite{Shiu}.

The delay features are influenced by two characteristic times: the damped Rabi oscillation period $\tau_{\rm R}=2\pi/\sqrt{|\Omega_c|^2-\Gamma^2/4}$ and the EIT group delay time $\tau_{\rm EIT}=\Gamma{\rm OD}/|\Omega_c|^2$~\cite{Wen1, Wen2}. The former is due to coherence oscillation caused by $\Omega_c$ during the conversion into anti-Stokes photons, while the latter results from the EIT slow light effect. Please note that both $\tau_{\rm R}$ and $\tau_{\rm EIT}$ are applicable only when the coupling field is resonant; they cannot be simply determined using these formulas under detuned conditions. As shown in Fig.~\ref{fig:slowlight}(a), where $\tau_{\rm EIT}=265$ ns is larger than $\tau_{\rm R}=192$ ns, the EIT slow light effect dominates, indicating the slow light regime, as evidenced by the delay tail. When tuning the coupling field frequency $\omega_c$ away from resonance, the central angular frequency of the generated anti-Stokes field shifts to satisfy the two-photon resonance condition $\bar{\omega}_{as}=\omega_{21}+\omega_c=\omega_{41}+\Delta_c$, where $\omega_{21}$ and $\omega_{41}$ represent the angular frequencies of the relevant atomic transitions. This two-photon resonance condition ensures high transmission of the anti-Stokes field when tuning the biphoton frequency. 

Figure~\ref{fig:slowlight}(b) presents the biphoton wavepacket for blue-detuning the resonant scenario to $\Delta_c=+1\Gamma$. In this case, the delay time decreases due to the reduction of the effective OD, thereby degrading the EIT-induced delay. However, this phenomenon is not observed in Fig.~\ref{fig:slowlight}(c), which shows the red-detuned $\Omega_c$ case with $\Delta_c=-1\Gamma$. This difference arises because detuning $\Omega_c$ from resonance makes the conversion of ground-state coherence into anti-Stokes photons by $\Omega_c$ take longer, thereby leading to an increase in $\tau_{\rm R}$. The difference between the blue-detuned and red-detuned $\Omega_c$ cases is due to the phase-mismatch effect. As the magnitude of the coupling detuning $\Delta_c$ continues to increase, the influence of $\tau_{\rm R}$ becomes more pronounced. This is evident from the increased delay shown in Figs.~\ref{fig:slowlight}(d)--\ref{fig:slowlight}(f). In addition, although varying $\Delta_d$ can frequency-tune the generated Stokes photons, changing $\Delta_d$ primarily affects the generation rate without significantly modifying the shape of the temporally correlated biphoton wavepacket. Therefore, we did not vary $\Delta_d$ in our experiment. For a more detailed theoretical analysis of $\tau_{\rm EIT}$ and $\tau_{\rm R}$, please refer to the supplemental material of our previous work~\cite{Shiu}.

%%%%%%%%%%%%%%%%%%%%%%%%%%%%%
\FigFour
%%%%%%%%%%%%%%%%%%%%%%%%%%%%%

%%%%%%%%%%%%%%%%%%%%%%%%%%%%%
\FigFive
%%%%%%%%%%%%%%%%%%%%%%%%%%%%%

Figure~\ref{fig:slowlight2}(a) illustrates the delay time $\tau_{\rm delay}$ of the biphoton wavepacket under various $\Delta_c$ values. The $\tau_{\rm delay}$ values obtained from the biphoton wavepackets in Fig.~\ref{fig:slowlight} are shown as circles in Fig.~\ref{fig:slowlight2}(a), where they are determined based on the relation $r_p^{-1}\int_0^{\tau_{\rm delay}}[R_{\rm C}(\tau)-R_{\rm C}(\infty)]d\tau=1-e^{-1}$.  These results agree well with the theoretical predictions indicated by the red solid curve. The black dotted curve represents the phase-mismatch-free case, where $\tau_{\rm delay}$ shows symmetric outcomes for red- and blue-detuned $\Omega_c$. Additionally, the blue dashed curve represents the scenario with phase mismatch twice as large as our experimental value, $\Delta kL=0.74\pi$. In this case, the asymmetry in $\tau_{\rm delay}$ for red- and blue-detuned $\Omega_c$ becomes more pronounced. Therefore, characterizing the $\tau_{\rm delay}$ curve as a function of $\Delta_c$ can serve as a method for determining the length of the atomic medium. Notably, phase mismatch has minimal impact on $\tau_{\rm delay}$ under resonant conditions. This explains why previous studies have rarely addressed the phase mismatch effect.

Figure \ref{fig:slowlight2}(b) demonstrates the variation in the biphoton pairing ratio $r_p$ across different values of coupling detuning $\Delta_c$. The introduction of $\Delta_c$ results in a decrease in $r_p$ due to its weakening effect on EIT-based stimulated FWM, which is essential for pairing the generated photons. When the generated photons cannot effectively interact with neighboring atoms, making stimulated four-wave mixing less likely, $r_p$ decreases. Additionally, the variations in the generation rate of biphotons $R_{\rm B}$ and the generation rate of temporally correlated photons $R_{\rm B}r_p$ are shown in Figs. \ref{fig:slowlight2}(c) and \ref{fig:slowlight2}(d), respectively. The experimental $R_{\rm B}$ values are determined by subtracting $R_{\rm env}$ from the experimental $R_{\rm C}$, where $R_{\rm env}$ is estimated based on optical leakage and dark counts from SPCMs. $R_{\rm B}$ represents the photon generation rate scattered in the designated direction, while $R_{\rm B} r_p$ specifically denotes the photon generation rate associated with the stimulated FWM process. Both decrease with increasing the magnitude of $\Delta_c$. These results indicate that tuning the biphoton frequency leads to changes in the wavepacket temporal profile, pairing ratio, and generation rate.

Despite the fact that introducing $\Delta_c$ leads to a decrease in the performance of the biphoton source, such as reducing $r_p$ and $R_{\rm B}$, this issue can be addressed by using a higher power coupling field. Specifically, a larger $\Omega_c$ can enhance the EIT-based stimulated FWM effect, making the biphoton pairing ratio and generation rate less susceptible to the influence of $\Delta_c$. As shown in Fig. \ref{fig:Rabi}, the biphoton wavepackets with a larger $\Omega_c=2\Gamma$ demonstrate this effect. At $\Delta_c=0\Gamma$, $\tau_{\rm R}=86$ ns is larger than $\tau_{\rm EIT}=66$ ns, placing the biphoton wavepacket in the damped Rabi oscillation regime, unlike the slow light regime shown in Fig. \ref{fig:slowlight}. Under ideal conditions, since the area of the temporally correlated biphoton wavepacket, or $r_p$, hardly changes with $\Omega_c$, this results in a short correlation time. Therefore, increasing $\Omega_c$ can ameliorate the decrease in the signal-to-background ratio (SBR) caused by tuning the biphoton frequency. Notably, the SBR of the biphoton wavepacket in Fig. \ref{fig:Rabi}(a) is 43, significantly surpassing the ratio of 13 observed in Fig. \ref{fig:slowlight}(a) when $\Omega_c=1\Gamma$.

Figure \ref{fig:Rabi2}(a) shows the variation of $\tau_{\rm delay}$ with the magnitude of $\Delta_c$. Unlike the trend observed in Fig. \ref{fig:slowlight2}(a), in the phase-mismatch-free scenario (black dotted curve), $\tau_{\rm delay}$ does not decrease with increasing $\Delta_c$. This behavior can be attributed to the dominance of damped Rabi oscillations in the biphoton wavepacket in Fig. \ref{fig:Rabi2}. Further increasing $\Delta_c$ causes $\tau_{\rm R}$ to increase and $\tau_{\rm EIT}$ to decrease. This explains why, in the damped Rabi oscillation regime, increasing $\Delta_c$ does not result in a decrease in the delay. 

Figures \ref{fig:Rabi2}(b) and \ref{fig:Rabi2}(c) showcase the variations in $r_p$ and $R_{\rm B}$ across different $\Delta_c$ values. While both $r_p$ and $R_{\rm B}$ exhibit a decline with increasing $\Delta_c$, this reduction is more gradual compared to the scenario with $\Omega_c=1\Gamma$. This phenomenon arises because increasing $\Omega_c$ enhances the interaction between photons and atoms, thereby mitigating the weakening effect caused by $\Delta_c$ and enhancing the efficacy of the EIT-based stimulated FWM process. The generation rate of correlated photon pairs, $R_{\rm B}r_p$, is shown in Fig. \ref{fig:Rabi2}(d). Unlike the scenario with $\Omega_c=1\Gamma$, the decay trend with increasing $\Delta_c$ is not prominent. This indicates that although tuning the biphoton frequency leads to a decrease in $R_{\rm B}$ and $r_p$, this trend can be mitigated by increasing $\Omega_c$. These values of $R_{\rm B}$, $r_p$, and $R_{\rm B}r_p$, which are not significantly reduced by the introduction of $\Delta_c$, indicate that biphoton performance can be maintained while tuning the biphoton frequency, demonstrating the tunability of the biphoton frequency. We emphasize that the current experimental results demonstrate a biphoton frequency tuning range of up to $3\Gamma$, or 18 MHz, which is sufficient for many frequency-sensitive systems. Based on theoretical predictions, if the required $|\Delta_c|$ increases by a factor of $n$, the biphoton performance (characterized by $R_{\rm B}$ and $r_p$) can be nearly preserved by scaling $\Omega_c$ proportionally to $|\Delta_c|$. For instance, increasing the coupling field Rabi frequency $\Omega_c$ from the current value of $2\Gamma$ to $20\Gamma$—a tenfold increase—would extend the tunable range to $|\Delta_c|=30\Gamma$ (180 MHz) while maintaining biphoton performance. This linear dependence of $|\Delta_c|$ on $\Omega_c$ highlights the feasibility of significantly expanding the tunable range by optimizing the coupling field power in practical implementations. These results reinforce the practicality of frequency-tunable biphoton generation using the double-$\Lambda$ scheme for diverse quantum technologies.

%%%%%%%%%%%%%%%%%%%%%%%%%%%%%%%%%%%%%%%%%%%%%%%%%%%%%%%%%%%%%%%%%%%%%%%%%%%%%%%%%%%%%%%%%%%%%%%%%%%%%
%%%%%%%%%%%%%%%%%%%%%%%%%%%%%%%%%%%%%%%%%%%%%%%%%%%%%%%%%%%%%%%%%%%%%%%%%%%%%%%%%%%%%%%%%%%%%%%%%%%%%

\section{Conclusions} \label{sec:Conclusion}

In summary, our experimental investigation of the pairing ratio through tuning biphoton frequency via backward SFWM in cold atoms reveals several insights. By manipulating the inherent EIT effect in the double-$\Lambda$ scheme, we demonstrate that introducing a detuned coupling field negatively impacts the efficiency of EIT-based stimulated FWM, thereby reducing the biphoton pairing ratio. However, this adverse effect can be mitigated by increasing the optical power of the incident coupling field. Moreover, we observe an asymmetry in the temporal profiles of biphotons when blue- or red-detuning their frequencies, attributed to the phase-mismatch effect. These observations not only enhance the understanding of the underlying mechanisms in biphoton generation but also suggest potential optimizations for applications in quantum communication and information processing.

%%%%%%%%%%%%%%%%%%%%%%%%%%%%%%%%%%%%%%%%%%%%%%%%%%%%%%%%%%%%%%%%%%%%%%%%%%%%%%%%%%%%%%%%%%%%%%%%%%%%%
%%%%%%%%%%%%%%%%%%%%%%%%%%%%%%%%%%%%%%%%%%%%%%%%%%%%%%%%%%%%%%%%%%%%%%%%%%%%%%%%%%%%%%%%%%%%%%%%%%%%%

\section*{ACKNOWLEDGEMENTS}

This work was supported by the National Science and Technology Council of Taiwan under Grants No. 111-2112-M-006-027, No. 112-2112-M-006-034, and No. 112-2119-M-007-007. We also acknowledge support from the Center for Quantum Science and Technology (CQST) within the framework of the Higher Education Sprout Project by the Ministry of Education (MOE) in Taiwan.

%%%%%%%%%%%%%%%%%%%%%%%%%%%%%%%%%%%%%%%%%%%%%%%%%%%%%%%%%%%%%%%%%%%%%%%%%%%%%%%%%%%%%%%%%%%%%%%%%%%%%
%%%%%%%%%%%%%%%%%%%%%%%%%%%%%%%%%%%%%%%%%%%%%%%%%%%%%%%%%%%%%%%%%%%%%%%%%%%%%%%%%%%%%%%%%%%%%%%%%%%%%

%%%%%%%%%%%%%%%%%%%%%%%%%%%%%%%%%%%%%%%%%%%%%%%%%%%%%%%%%%%%%%%%%%%%%%%%%%%%%%%%%%%%%%%%%%%%%%%%%%%%%
%%%%%%%%%%%%%%%%%%%%%%%%%%%%%%%%%%%%%%%%%%%%%%%%%%%%%%%%%%%%%%%%%%%%%%%%%%%%%%%%%%%%%%%%%%%%%%%%%%%%%

\end{document}